# Skyrmion Tubes as Magnonic Waveguides

Xiangjun Xing,*,† Yan Zhou,*,‡ and H. B. Braun*,§

†*School of Physics and Optoelectronic Engineering, Guangdong University of Technology, Guangzhou 510006, China*

‡*School of Science and Engineering, The Chinese University of Hong Kong, Shenzhen, 518172, China*

§*School of Physics, University College Dublin, Dublin 4, Ireland*

Various latest experiments have proven the theoretical prediction that domain walls in planar magnetic structures can channel spin waves as outstanding magnonic waveguides, establishing a superb platform for building magnonic devices. Recently, three-dimensional nanomagnetism has been boosted up and become a significant branch of magnetism, because three-dimensional magnetic structures expose a lot of emerging physics hidden behind planar ones and will inevitably provide broader room for device engineering. Skyrmions and antiSkyrmions, as natural three-dimensional magnetic configurations, are not considered yet in the context of spin-wave channeling and steering. Here, we show that skyrmion tubes can act as nonplanar magnonic waveguides if excited suitably. An isolated skyrmion tube in a magnetic nanoprism induces spatially separate internal and edge channels of spin waves; the internal channel has a narrower energy gap, compared to the edge channel, and accordingly can transmit signals at lower frequencies. Additionally, we verify that those spin-wave beams along magnetic nanoprism are restricted to the regions of potential wells. Transmission of spin-wave signals in such waveguides results from the coherent propagation of locally driven eigenmodes of skyrmions, *i.e.*, the breathing and rotational modes. Finally, we find that spin waves along the internal channels are less susceptible to magnetic field than those along the edge channels. Our work will open a new arena for spin-wave manipulation and help bridge skyrmionics and magnonics.

---

*Email: xjxing@gdut.edu.cn; zhouyan@cuhk.edu.cn; beni.braun@ucd.ie

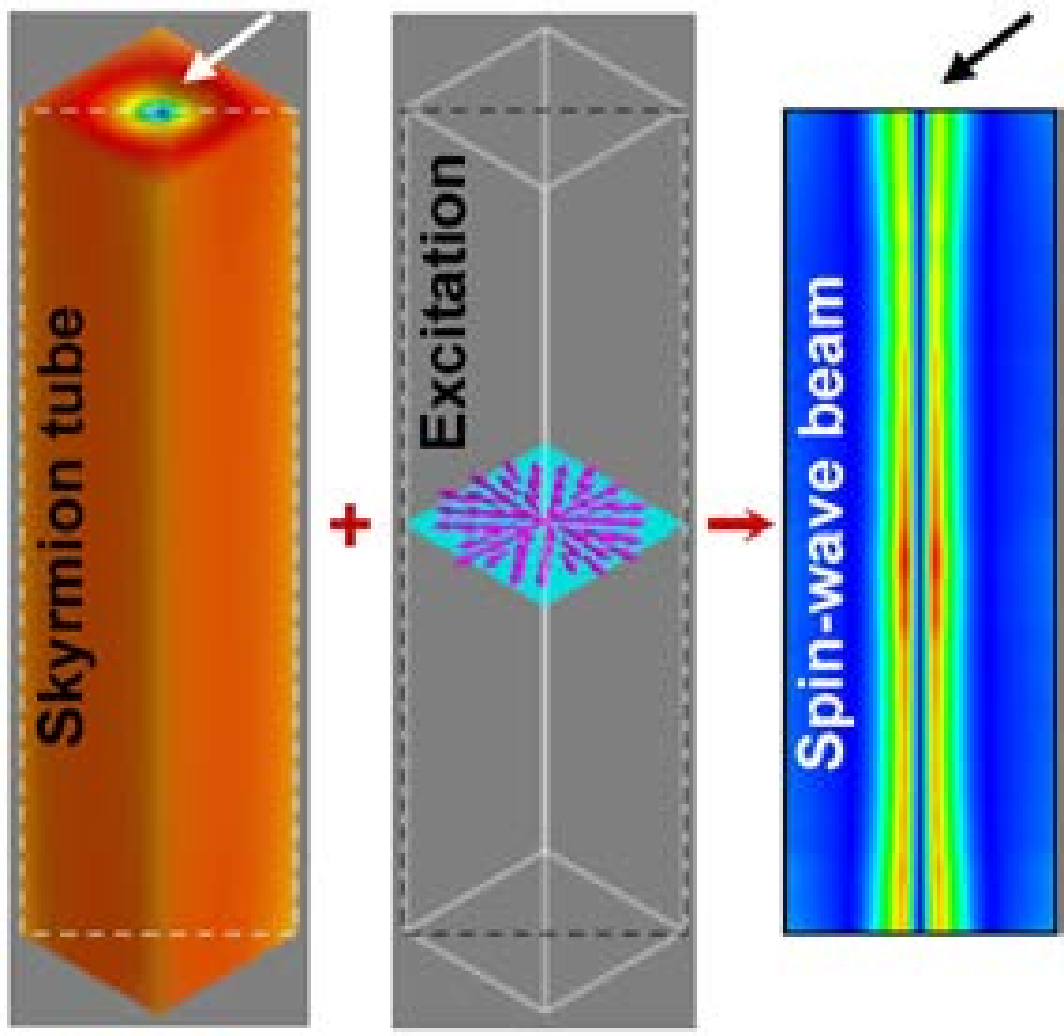

Skyrmion tube + Excitation → Spin-wave beam

Magnonic devices hold great promise to complement or even outperform conventional electronic devices in beyond-CMOS era (1-4), especially when implementing non-Boolean computations (5) for special tasks, *e.g.*, pattern recognition (6,7). The benefits offered by magnonic devices crucially rely on a key building block — magnonic waveguides, which govern the propagation of spin-wave signals (3,4). Recently, magnetic domain walls were theoretically proposed as magnonic waveguides (8,9); this idea has been verified by myriad latest experiments (10,11), establishing a prominent platform for constructing magnonic devices (12,13). Meanwhile, three-dimensional (3D) nanomagnetism has been actively pursued and become an important direction in magnetism (14). On the one hand, a lot of emergent physics in the frame of 3D nanomagnetism has been identified, *e.g.*, the motion of domain walls in nanotubes and circular nanowires can avoid undergoing any instability such as the Walker breakdown (15); on the other hand, 3D nanomagnetism will enrich the toolset for device engineering (16). The widely used 3D NAND flash memories in present-day adopt a stack structure of planar elements (4). In fact, in the pioneering work of domain-wall racetrack memory, Parkin *et al.* have conceived a nonplanar version of that memory to realize an ultrahigh storage density by making use of the third dimension (17).

Magnetic skyrmions provide the benefit to utilize the third dimension, since, in several material systems, they appear in the form of tubes aligned perpendicular to plane (18-23). During the past few years, studies on magnetic skyrmions mainly focused on exploring a wealth of properties of skyrmions in bulk crystals and thin plates of chiral magnets as well as ultrathin magnetic multilayers (24-26), and on developing a variety of related devices that use skyrmions as information carriers (26-29). However, no research has attempted to exploit the possibility to employ skyrmions as delivery channels of signals. In our view, magnetic skyrmions are good magnonic waveguides, allowing for construction of advanced nonplanar magnonic devices.

Oscillation of local magnetic moments can travel and thereby produce spin waves in a magnet, if the temporal and spatial structure of local magnetic excitations match the relevant parameters of magnetic system. The ability of magnetic domain walls to act as magnonic waveguides originates from the coherent propagation of locally excited Winter modes of domain walls (8). A domain wall creates a potential well, which guides spin waves along the domain wall (9). Similarly, the eigenexcitations of a magnetic skyrmion include breathing and rotational modes (30). Under suitable conditions, coherent transmission of locally excited breathing and rotational modes should be possible via the generation of propagating spin waves along a skyrmion tube. In this way, a skyrmion tube becomes a magnonic waveguide.

In this work, we examine the excitation and transmission of spin waves in a magnetic square nanoprism containing the spin configuration of a skyrmion tube and compare it to the single-domain case. The magnon dispersion relations reveal that traveling spin waves exist in the magnetic square prism, irrespective of the spin configuration. However, when the prism hosts a skyrmion tube, an extra dispersion branch appears that has an ultralow energy gap. A Fourier analysis of the dynamic magnetization over the entire space, identifies the emergent dispersion branch, as a spin-wave mode that is transversely restricted to an area around the core of the skyrmion tube, but propagates along the axis of the tube. Namely, the magnetic square prism containing a skyrmion tube becomes a self-organized magnonic waveguide with an internal channel, resembling an optic fiber (31). Furthermore, it is seen that spin waves propagating along the internal channel are less sensitive to a variation in the applied field than those travelling along the edge channel. Our results mark a new direction for technological applications of magnetic skyrmions in magnonics.

Our goal is to determine the wave-guiding characteristics of magnetic skyrmion tubes. In order to make the computations easily accessible, we chose a square prism with width $w$ = 100 nm and length $l$ = 500 nm which can accommodate an isolated skyrmion tube. A sketch of our studied sample is shown in Fig. 1, and the structure can support various magnetization states, depending on material parameters and magnetic history. The type of the skyrmion, *i.e.*, Skyrmion or antiSkyrmion (Skyrmions and antiSkyrmions are jointly called skyrmions. Skyrmions have topological charge $Q$ = 1, while antiSkyrmions refer to textures with $Q$ = -1) existing in the prism relies on the property of Dzyaloshinskii-Moriya interaction (DMI) (32,33), which is, in turn, determined by the crystal symmetry of magnetic materials (34). Figs. 1(a) and 1(b) illustrate configurations consisting of an antiSkyrmion tube and a quasiuniform single-domain state in a sample with D2d-type DMI (23,35,36). Figs. 1(c) and 1(d) show the Skyrmion tube and quasiuniform single domain in a sample with bulk-type DMI (18,35,36). Also a top view of the spin configurations across all the cross-sectional planes is shown. It is worth noting that our simulated skyrmion's structure replicates the surface reconstruction of skyrmions that was viewed as a universal phenomenon arising from the broken translational symmetry at an interface (37). Fig. 1(e) depicts the spatial profile $\mathbf{S}(x, y)$ of the time dependent excitation field $\mathbf{h}(x, y; t)$ used for the local injection of magnons.

We performed micromagnetic simulations to track the dynamics of magnetic moments in the prism. A local excitation is created by an embedded antenna (c.f. supplementary Fig. S1), and the magnetization dynamics is described by the modified Landau-Lifshitz-Gilbert equations including DMI,

$$\partial_t \mathbf{m} = -\gamma (\mathbf{m} \times \mathbf{H}_{eff}) + \alpha (\mathbf{m} \times \partial_t \mathbf{m}),$$

where $\mathbf{m} = \mathbf{M}/M_s$ is a space and time dependent unit vector representing the magnetization $\mathbf{M} = \mathbf{M}(x,$

$y$, $z$; $t$) normalized by its saturation value $M_s$, and $\mathbf{H}_{eff} = -\,(1/\mu_0)\delta E/\delta \mathbf{M}$ is the effective field with $\mu_0$ denoting the vacuum permeability. The total energy $E = E_d + E_x + E_u + E_{DM} + E_Z$ is the sum of magnetostatic, exchange, anisotropy, DMI and Zeeman energy terms.

The open-source code OOMMF (38) was used to carry out all the simulations. We searched for the equilibrium magnetization states of the prism by relaxing it from a numerically conjectured skyrmion-like state over a range of values of $H_{bias}$ (the static biasing field, always along $+z$), $K_u$ (the perpendicular magnetocrystalline anisotropy), and $D$ (the DMI strength), and identified the parameter window in which a single skyrmion tube can exist as a (meta)stable spin configuration. Three sets of simulations were made to investigate the dependence of equilibrium spin configuration on material parameters. In each simulation set, one of the three quantities $D$, $K_u$, and $H_{bias}$ is chosen to be free, while the others stay fixed. Here, we specially chose a shorter prism with a size of $100{\times}100{\times}100$ nm$^3$ and discretized it using $1{\times}1{\times}1$ nm$^3$ cells for better numerical accuracy. In each individual calculation, we start from a skyrmion-like spin texture in the prism and then let the entire system freely relax, till a specified stopping criterion, *i.e.*, $\max\{|\mathbf{m}{\times}(\mathbf{H}_{eff}{\times}\mathbf{m})|\} = 0.01$ A/m, is met. It is seen that a single skyrmion tube can exist in the prism at equilibrium over a broad interval in the parameter space.

In dynamic simulations, we did not directly include an antenna, but simply adopted an excitation field with a spatial profile as specified in Fig. 1(e), to mimic the real field arising from an antenna (39). The prism was divided into a regular array of meshes of $2{\times}2{\times}2$ nm$^3$ to attain a tradeoff between computation speed and numerical accuracy. The Gilbert damping was set to $\alpha = 0.01$ (40) throughout the prism except at the absorbing boundary region, where its value is enhanced to 0.5 or higher (41). For a given type of DMI, the considered magnetic background for the magnons is either a single skyrmion-tube or a quasiuniform single-domain state. The excitation field $\boldsymbol{h}(x, y; t) = H_0 \cdot \boldsymbol{S}(x, y) \cdot P(t)$, is applied locally to the middle cross-sectional plane of the prism, and has a spatial profile $\boldsymbol{S}(x, y) = (x\hat{\boldsymbol{e}}_x + y\hat{\boldsymbol{e}}_y)/\sqrt{x^2 + y^2}$ [c.f. Fig. 1(e)] with a temporal profile $P(t) = \sin[2\pi f_c(t{-}t_0)]/[2\pi f_c(t{-}t_0)]$, where $H_0 = 20$ kOe and $f_c = 200$ GHz are the amplitude and cutoff frequency of the *sinc* field, respectively. To derive magnon dispersion relations and mode patterns, we evaluated the magnetization $\boldsymbol{M}(x, y, z; t)$ of the prism every $\Delta T = 2.475$ ps within a fixed duration after excitation, leading to the excitation frequency bandwidth $f_i = 1/(2\Delta T) = 202.0$ GHz, so that the condition $f_c < f_i$ to prevent aliasing is satisfied (42). The time sequence of magnetization was analyzed via Fourier transform with the SEMARGL software following the procedure described in Ref. (42). The static magnetic background was subtracted from the transient states and thus did not enter the frequency-domain analyses.

The presented results are based on the following material parameters, unless otherwise specified: saturation magnetization $M_s = 445$ kAm$^{-1}$, exchange stiffness $A = 120$ pJm$^{-1}$, uniaxial anisotropy constant $K_u = 0.3$ MJm$^{-3}$, DMI constant $D = 6.0$ mJm$^{-2}$, and bias field $H_{bias} = 5$ kOe. These parameters correspond to the experimental values reported for the inverse-Heusler compound Mn-Pt-Sn (23), and define two length scales: the exchange length $\Lambda = \sqrt{2A/\mu_0 M_s^2} \approx 31.1$ nm (the maximum length at which the short-range exchange interaction can maintain a uniform magnetization) and the domain wall thickness $\delta = \sqrt{A/K_u} \approx 20.0$ nm. For both static and dynamic simulations, the used mesh size is much smaller than the size of the magnetic features in order to prevent numerical artefacts (43).

To clarify the wave-guiding characteristics, we inspect the magnon dispersion along five separate straight lines (marked as 1–5 in the inset to Fig. 2), which all run parallel to the long axis of the prism. Five sets of dispersion relations for the antiSkyrmion tube and quasiuniform single domain are shown in Figs. 2(a) and 2(b), respectively. Clearly, there are propagating spin waves along the square prism, regardless of the assumed static spin configuration. For the antiSkyrmion tube, the lines #1–5 display diverse dispersion characters. Going from the periphery to the interior of the cross-sectional plane, a low-lying dispersion branch emerges and gradually dominates the original branches. However, for the quasiuniform single domain, from lines #1 to #5, the entire set of dispersion curves barely change except for some slight fluctuation in the Fourier amplitude. That is, new propagating magnon modes can be brought into the prism, when the quasiuniform single domain is transformed into the antiSkyrmion tube. These nascent spin waves travel along the inner line (#3–5) and exhibit a reduced band gap, similar to propagating spin waves along domain walls in a magnonic fiber waveguide (9).

To uncover the nature of excitations in the prism, we show the mode propagation patterns for excitations at chosen frequencies; some of them are shown in Fig. 3. In each panel, we represent the Fourier amplitude and phase profiles of excitations on three rectangular cut-planes ($x = w/2$; $y = w/2$; $y = x$) and three cross-sectional planes ($z = l/4$, $l/2$, $3l/4$), to resolve the mode structure.

Figures 3(a) and 3(b) depict the magnon mode patterns, at 19.76 GHz, for the antiSkyrmion tube and quasiuniform single domain, respectively. For the former, the Fourier amplitude profiles on the rectangular cut-planes indicate that spin-wave beams are restricted to the interior of the prism and surround the antiSkyrmion core line. The pronounced phase accumulation along the $z$-axis, as shown in the Fourier phase profiles, illustrates the propagating character of such excitations along the internal channel. Spatial localization of the channel around the antiSkyrmion core is perfectly reflected in the amplitude patterns from the cross-sectional planes. The mode patterns on the cross-sectional planes are reminiscent of two categories of eigenexcitations of a skyrmion. Actually, the patterns on the cross-sectional planes at $z = l/2$ and $l/4$ ($3l/4$), can be assigned to the rotational and breathing modes of a skyrmion, respectively.

For the quasiuniform single-domain state, the frequency 19.76 GHz falls into the forbidden band; therefore, spin waves at this frequency will attenuate immediately in space after emission from the excitation region (see Fig. 3(b), the amplitude patterns on the rectangular cut-planes) and form chaotic mode patterns on the two off-center cross-sectional planes. Nevertheless, once the excitation frequency surpasses a threshold value (~32 GHz), such edge spin waves can escape from the magnon injection region and freely travel along the prism. Fig. 3(d) shows the mode patterns above the quasiuniform single-domain state for the spin waves at 34.58 GHz, at which the spin-wave wavelength almost occupies the entire length of the prism. The beating patterns on the cross-sectional planes reveal the mode structure of the edge spin waves.

For the antiSkyrmion tube, continued increase of the excitation frequency will lead to spin waves of shorter wavelengths [c.f. Figs. 3(a), 3(c), S2(a), and S2(c)], and once the frequency is above the threshold value, edge spin waves will be generated in the prism, in tandem with the interior spin-wave beams along the internal channel [Figs. 3(c), S2(a), and S2(c)]. The coexisting internal and edge spin waves, at the same temporal frequency, have slightly different spatial periods, as shown in the phase profiles from the rectangular cut-planes. This is also visible from the dispersion relations [see Fig. 2(a), #4]. The Fourier amplitude profiles on the cross-sectional planes [Fig. 3(c)] offer an additional proof for the concurrence of edge and internal spin waves.

Much information can be obtained if one compares those transverse mode patterns at various individual frequencies. First, the excitations in the antenna region always have a mode structure characteristic of the rotational mode, but after diffusion away from the source, they may retain rotational character [Figs. 3(c), S2(a), and S2(c)] or convert to the breathing mode [Fig. 3(a)]. Second, the original rotational mode has two alternative senses of rotation (30): counterclockwise (CCW) for some excitation frequencies [Figs. 3(a), 3(c), and S2(c)] and clockwise (CW) for the other [Fig. S2(a)]. In the latter case, we observe a structural transition of the rotational mode from CW to CCW during its propagation. Third, the original rotational modes always manifest an amplitude profile with fourfold symmetry, but when coupled into the channel they could assume a twofold symmetry, as shown in Figs. S2(a) and S2(c).

Now, let us turn to the analysis of the results (Figs. 4, 5, and S3) for the Skyrmion tube and quasiuniform single-domain state rendered by the bulk-type DMI. Through a direct comparison, we see that the primary findings, for the antiSkyrmion tube and quasiuniform single domain governed by the D2d-type DMI, remain valid in the current case. Remarkably, there also exist propagating spin waves in the square prism, for both the Skyrmion tube and quasiuniform single domain, and additionally, writing an isolated Skyrmion tube into the prism can still give rise to an internal spin-wave channel inside the prism. Except for the stated analogies, there are also some critical differences between Skyrmion and antiSkyrmion tubes. For instance, internal and edge spin waves, for the Skyrmion-tube state, sharply differ in their spectral amplitudes, despite identical excitation conditions. As a result, the dim edge mode is hardly visible in the Fourier amplitude maps when the bright internal mode predominates. Moreover, throughout the entire frequency range, we do not observe the occurrence of rotational modes; the excitation and propagation of breathing modes is responsible for the formation of internal channel in the prism bearing a Skyrmion tube. Finally, while for the antiSkyrmion tube and quasiuniform single domain under the D2d-type DMI all spin waves along the prism are reciprocal, for the Skyrmion tube and quasiuniform single domain under the bulk-type DMI, all spin waves are nonreciprocal (44), as revealed by the dispersion data and indicated in the mode patterns. The difference in the reciprocity of spin waves has its origin in the different forms of DMI required for stabilizing various types of skyrmions.

For both types of skyrmion tubes, we use the same excitation field with cylindrical symmetry. Then, the unique symmetry of the magnetic background determines the symmetry of the allowed eigenmodes. The magnetization configuration in an antiSkyrmion tube breaks the cylindrical symmetry and generates a quadrupolar moment of magnetostatic charges (23,45), whereas that in a Skyrmion tube preserves the cylindrical symmetry. This structural difference between the Skyrmion and antiSkyrmion tubes is the reason why the rotational modes are not activated for the Skyrmion tube. Simply lowering the symmetry of the excitation field, $e.g.$, redirecting it along the $x$-axis, would enable the excitation of rotational modes on the Skyrmion tube.

In the preceding sections, we analyzed the results in the language of eigenmodes of magnetic skyrmions (30). Indeed, the energy-well picture (46) may provide a more direct understanding of spin-wave channeling along skyrmion tubes. Figure 6 shows the internal-field landscapes for the magnetic skyrmions and quasiuniform single domains, suggesting the existence of potential wells. For the skyrmion tubes, the energy well surrounds the skyrmion core where an exceedingly high energy barrier arises. In contrast, for the quasiuniform single domain, the energy well appears at the boundary region. Actually, the internal field of the skyrmion tubes also involves an edge potential well, which is nevertheless not resolved in this figure because of a small depth.

Once excited, an energy well will support localized modes, and then propagate them if the well is adequately extended in a certain direction (9). This lays the foundation for guiding spin waves through the skyrmion tubes. It can also explain why the internal channels exhibit a much smaller band gap considering the fact that the central potential well is deeper than that at the edge .

So far, both antiSkyrmion and Skyrmion tubes have not yet been stabilized experimentally without applying a perpendicular biasing field. Generally, a skyrmion tube can exist in a range of magnetic fields (18-23). Thus, one should investigate how the perpendicular field affects spin waves flowing along a skyrmion tube. To this end, we compare the dispersion relations of spin waves for different biasing fields. For each type of skyrmion tube, three field values are selected: $\mu_0 H_{bias}$ = 0.4, 0.5, and 0.6 T for the antiSkyrmion tube and 0.5, 0.6, and 0.7 T for the Skyrmion tube. If $\Delta f_b$ represents the net increase in the bottom frequency $f_b$ of a dispersion branch that follows an increase $\mu_0 \Delta H_{bias}$ in the biasing field, a proportionality factor, $\eta = \Delta f_b / \mu_0 \Delta H_{bias}$, can be defined for each dispersion branch. Figure 7 presents a comparison, from which we can identify several facts. For the antiSkyrmion tube, all dispersion curves shift upward along the frequency axis as the bias field increases, *i.e.*, $\eta$ is always positive, and the modes with higher frequencies are more sensitive to variation in the biasing field strength. For the Skyrmion tube, however, while the lower-frequency mode shifts upward along the frequency axis ($\eta > 0$), the higher-frequency mode shifts downward ($\eta < 0$).

We would like to make some additional notes regarding the dispersion maps as well as the propagation patterns. Turning back to Figs. 2 and 4, we are aware that each panel in these figures contains multiple dispersion branches, implying that, for both the internal and edge channels, not only the lowest-order (fundamental) mode is excited by the excitation field, but also the higher modes are excited, causing the copropagation of multimodes with different order numbers (47). These coexisting multiple modes superpose and interfere with each other, which is the reason why periodic amplitude modulation frequently happens to the propagation patterns (47). Identification of the order number and mode structure of each mode requires elaborate data-fitting work (48) and is beyond the scope of the current study.

Moreover, the propagation patterns for the Skyrmion tube look cleaner than that for the antiSkyrmion tube. This might be ascribed to the sharp contrast between the amplitudes of the edge and internal spin waves with the Skyrmion tube, and the fact that, for the Skyrmion tube, excitation of the rotational modes is prohibited and only the breathing mode is permitted suppressing the number of allowed modes.

Because of the decreased symmetry of the magnetization distribution, the antiSkyrmion tube exhibits more complex mode excitations compared to the Skyrmion tube, which makes it more difficult to understand the behavior of spin waves in the antiSkyrmion tube, in particular the mechanism for the demonstrated mode conversion. More systematic investigations are needed to elucidate the full detail. In particular, other forms of excitation fields with distinctive symmetries should be employed to activate more available modes. In spite of the absorptive boundary condition, the spin-wave propagation patterns exhibit the character of standing waves, complicating the analysis of mode structures. In this process, part of the spin-wave energy is reflected back from the physical boundary and interferes with the forward-propagating spin waves. Usage of an extremely long prism can suppress this reflection effect but will sharply increase the computation time. In simulations of the Skyrmion tube, we did not change the material parameters from those used for the antiSkyrmion tube, apart from changing the Lifshitz invariant (34,35) to yield a bulk-type DMI. Consequently, the material parameters (23) used for the Skyrmion tube are not representative of typical materials in which Skyrmions are commonly found. This fact does not restrict the validity of the relevant results, because the qualitative aspect rather than the numeric precision of the findings is at the core of the present work exploiting the feasibility of channeling spin waves through skyrmion tubes.

Up to now, magnetic Skyrmions have been observed experimentally in various material systems including the B20 alloys (18), multiferroic $Cu_2OSeO_3$ (19), $\beta$-Mn-type Co-Zn-Mn alloys (20), polar magnets $GaV_4S_8$ (21) and $VOSe_2O_5$ (22), as well as ultrathin magnetic multilayers (25,26), whereas only the inverse-Heusler alloys Mn-Pt-Sn have been experimentally identified as a host of magnetic antiSkyrmions (23). Magnetic multilayers are not expected to support Skyrmion tubes that allow the propagation of coherent spin waves. FeGe and Co-Zn-Mn alloys allow the formation of Skyrmions above room temperature at lower fields (20) than that for the other aforementioned materials; the spectroscopy of propagating spin waves in these alloys has recently been used to infer the relevant DMI (49). AntiSkyrmions were first observed near room temperature in an inverse-Heusler Mn-Pt-Sn compound, a member of the Heusler family (50) that usually combines high spin polarization and a high Curie temperature with low Gilbert damping. In fact, spin-wave transport in such alloys has been investigated for about ten years (51). Therefore, among the materials listed above, the Mn-Pt-Sn and Co-Zn-Mn alloys seem best suited for near-term experimental implementation and long-term technological applications of magnonic waveguides based on skyrmion tubes.

Using skyrmion tubes to guide spin waves permits constructing multichannel magnonic waveguides from an extended magnetic sample in which skyrmion tubes can occur as an array or a lattice. Such a waveguide no longer requires patterning a sample into restricted geometry which greatly simplifies the fabrication procedure. Interestingly, directional emission of spin waves (52) in a bulk sample from a point source will be possible if magnetic bobbers (53) were utilized as magnonic waveguides. Intriguingly, a magnonic multiplexer (13,54) can even be realized in a bulk crystal of chiral magnets that contains a merging pair of Skyrmion tubes (55) wherein a spin-wave beam will split into two separate beams at the point of coalescence.

Vertical stacking of strip-shaped magnonic waveguides can yield nonplanar magnonic devices, which take advantage of the third dimension to save the chip area and thus would be desirable in practical applications. In such a stack structure, efficient signal delivery though the junction to all the involved planar elements and communication between any two elements are not easily attainable and anticipated to be fulfilled by magnetizing the entire stack into the out-of-plane direction and then injecting a skyrmion tube into the junction. Here, the introduction of the skyrmion tube, as a signal bus, is expected to enhance the transmission of spin waves.

The skyrmion-based magnonic waveguides share some analogies with their domain wall-based counterparts (8-13): Both of them originate from the coherent propagation of locally excited eigenmodes and are confined to the potential well associated with the static spin configuration. However, spin-wave beams inside them show striking difference: the beams along domain walls are "solid" and tied to the surface of the magnetic layer; those beams along skyrmion tubes are "hollow" and remote from the surface of the magnetic prism. Thus, it would be more challenging to detect the internal spin waves in the skyrmion-based magnonic waveguides.

In summary, we demonstrate through micromagnetic simulations that Skyrmion and antiSkyrmion tubes can function as nonplanar magnonic waveguides. This theoretical proposal of a novel type of spin-wave channeling mechanism, enables the creation of nanometer-scale spin-wave beams in extended chiral magnets without requiring hard structuring, and opens an alternative route toward nonplanar 3D magnonic devices.

**ASSOCIATED CONTENT**

**Supporting Information**

The schematic diagram of a thin toroidal antenna generating the excitation field and supplemental figures showing magnon propagation patterns and mode profiles at several additional frequencies.

## AUTHOR INFORMATION

**Corresponding Authors**

*E-mail: xjxing@gdut.edu.cn.

*E-mail: zhouyan@cuhk.edu.cn.

*E-mail: beni.braun@ucd.ie.

**Author Contributions**

X.J.X. initiated the study and performed micromagnetic simulations. Y.Z. coordinated the project. All authors contributed to the analysis of the results and wrote the manuscript.

**Notes**

The authors declare no competing financial interest.

## ACKNOWLEGMENTS

X.J.X. acknowledges the support from the National Natural Science Foundation of China (Grants No. 11774069 and No. 61675050). Y.Z. acknowledges the support by the National Natural Science Foundation of China (Grant No. 1157040329) and Shenzhen Fundamental Research Fund under Grant No. JCYJ20160331164412545.

## REFERENCES

(1) Kruglyak, V. V.; Demokritov, S. O.; Grundler, D. Magnonics. *J. Phys. D: Appl. Phys.* **2010**, 43 (26), 264001.

(2) Chumak, A.; Vasyuchka, V.; Serga, A.; Hillebrands, B. Magnon Spintronics. *Nat. Phys.* **2015**, 11 (6), 453–461.

(3) Lenk, B.; Ulrichs, H.; Garbs, F.; Munzenberg, M. The Building Blocks of Magnonics. *Phys. Rep.* **2011**, 507 (4–5), 107–136.

(4) The International Roadmap for Devices and Systems, 2017 Edition, "Beyond CMOS", https://irds.ieee.org/roadmap-2017

(5) Khitun, A. Magnonic Holographic Devices for Special Type Data Processing. *J. Appl. Phys.* **2013**, 113 (16), 164503.

(6) Kozhevnikov, A.; Gertz, F.; Dudko, G.; Filimonov, Y.; Khitun, A. Pattern Recognition with Magnonic Holographic Memory Device. *Appl. Phys. Lett.* **2015**, 106 (14), 142409.

(7) Romera, M.; Talatchian, P.; Tsunegi, S.; Araujo, F. Abreu; Cros, V.; Bortolotti, P.; Trastoy, J.; Yakushiji, K.; Fukushima, A.; Kubota, H.; Yuasa, S.; Ernoult, M.; Vodenicarevic, D.; Hirtzlin, T.; Locatelli, N.; Querlioz, D.; Grollier, J. Vowel Recognition with Four Coupled Spin-Torque Nano-Oscillators. *Nature* **2018**, 563 (7730), 230–234.

(8) Garcia-Sanchez, F.; Borys, P.; Soucaille, R.; Adam, J.-P.; Stamps, R. L.; Kim, J.-V. Narrow Magnonic Waveguides Based on Domain Walls. *Phys. Rev. Lett.* **2015**, 114 (24), 247206.

(9) Xing X.; Zhou, Y. Fiber Optics for Spin Waves. *NPG Asia Mater.* **2016**, 8, e246.

(10) Wagner, K.; Kákay, A.; Schultheiss, K.; Henschke, A.; Sebastian, T.; Schultheiss, H. Magnetic Domain Walls as Reconfigurable Spin-Wave Nanochannels. *Nat. Nanotechnol.* **2016**, 11 (5), 432–436.

(11) Albisetti, E.; Petti, D.; Sala, G.; Silvani, R.; Tacchi, S.; Finizio, S.; Wintz, S.; Calò, A.; Zheng, X.; Raabe, J.; Riedo, E.; Bertacco, R. Nanoscale Spin-Wave Circuits Based on Engineered Reconfigurable Spin-Textures. *Commun. Phys.* **2018**, 1, 56.

(12) Lan, J.; Yu, W.; Wu, R.; Xiao, J. Spin-Wave Diode. *Phys. Rev. X* **2015**, 5 (4), 041049.

(13) Xing, X.; Pong, P. W. T.; Åkerman, J.; Zhou, Y. Paving Spin-Wave Fibers in Magnonic Nanocircuits Using Spin-Orbit Torque. *Phys. Rev. Appl.* **2017**, 7 (5), 054016.

(14) Fernández-Pacheco, A.; Streubel, R.; Fruchart, O.; Hertel, R.; Fischer, P.; Cowburn, R. P. Three-Dimensional Nanomagnetism. *Nat. Commun.* **2017**, 8, 15756.

(15) Hertel, R. Ultrafast Domain Wall Dynamics in Magnetic Nanotubes and Nanowires. *J. Phys. Condens. Matter* **2016**, 28 (48), 483002.

(16) Lavrijsen, R.; Lee, J. H.; Fernandez-Pacheco, A.; Petit, D.; Mansell, R.; Cowburn, R. P. Magnetic Ratchet for Three-Dimensional Spintronic Memory and Logic. *Nature* **2013**, 493 (7434), 647–650.

(17) Parkin, S. S. P.; Masamitsu, H.; Thomas, L. Magnetic Domain-Wall Racetrack Memory. *Science* **2008**, 320 (5873), 190–194.

(18) Muhlbauer, S.; Binz, B.; Jonietz, F.; Pfleiderer, C.; Rosch, A.; Neubauer, A.; Georgii, R.; Boni, P. Skyrmion Lattice in A Chiral Magnet. *Science* **2009**, 323 (5916), 915–919.

(19) Seki, S.; Yu, X. Z.; Ishiwata, S.; Tokura, Y. Observation of Skyrmions in a Multiferroic Material. *Science* **2012**,

336 (6078), 198–201.

(20) Tokunaga, Y.; Yu, X. Z.; White, J. S.; Ronnow, H. M.; Morikawa, D.; Taguchi, Y.; Tokura, Y. A New Class of Chiral Materials Hosting Magnetic Skyrmions beyond Room Temperature. *Nat. Commun.* **2015**, 6, 7638.

(21) Kezsmarki, I.; Bordacs, S.; Milde, P.; Neuber, E.; Eng, L. M.; White, J. S.; Ronnow, H. M.; Dewhurst, C. D.; Mochizuki, M.; Yanai, K.; Nakamura, H.; Ehlers, D.; Tsurkan, V.; Loidl, A. Néel-Type Skyrmion Lattice with Confined Orientation in the Polar Magnetic Semiconductor $GaV_4S_8$. *Nat. Mater.* **2015**, 14 (11), 1116–1122.

(22) Kurumaji, T.; Nakajima, T.; Ukleev, V.; Feoktystov, A.; Arima, T.; Kakurai, K.; Tokura, Y. Néel-Type Skyrmion Lattice in the Tetragonal Polar Magnet $VOSe_2O_5$. *Phys. Rev. Lett.* **2017**, 119 (23), 237201.

(23) Nayak, A. K.; Kumar, V.; Ma, T.; Werner, P.; Pippel, E.; Sahoo, R.; Damay, F.; Rößler, U. K.; Felser, C.; Parkin, S. S. P. Magnetic Antiskyrmions above Room Temperature in Tetragonal Heusler Materials. *Nature* **2017**, 548 (7669), 561–566.

(24) Nagaosa N.; Tokura, Y. Topological Properties and Dynamics of Magnetic Skyrmions. *Nat. Nanotechnol.* **2013**, 8 (12), 899–911.

(25) Wiesendanger, R. Nanoscale Magnetic Skyrmions in Metallic Films and Multilayers: A New Twist for Spintronics. *Nat. Rev. Mater.* **2016**, 1 (7), 16044.

(26) Fert, A.; Reyren, N.; Cros, V. Magnetic Skyrmions: Advances in Physics and Potential Applications. *Nat. Rev. Mater.* **2017**, 2 (7), 15.

(27) Tomasello, R.; Martinez, E.; Zivieri, R.; Torres, L.; Carpentieri, M.; Finocchio, G. A Strategy for the Design of Skyrmion Racetrack Memories. *Sci. Rep.* **2014**, 4, 6784.

(28) Zhang, X.; Ezawa, M.; Zhou, Y. Magnetic Skyrmion Logic Gates: Conversion, Duplication and Merging of Skyrmions. *Sci. Rep.* **2015**, 5, 9400.

(29) Zhang, X.; Zhou, Y.; Ezawa, M.; Zhao, G. P.; Zhao, W. Magnetic Skyrmion Transistor: Skyrmion Motion in a Voltage-Gated Nanotrack. *Sci. Rep.* **2015**, 5, 11369.

(30) Mochizuki, M. Spin-Wave Modes and Their Intense Excitation Effects in Skyrmion Crystals. *Phys. Rev. Lett.* **2012**, 108 (1), 017601.

(31) Kao, C. K. Sand from Centuries Past: Send Future Voices Fast, *Nobel Lecture*, https://www.nobelprize.org/uploads/2018/06/kao_lecture.pdf

(32) Dzyaloshinskii, I. E. Thermodynamic Theory of 'Weak' Ferromagnetism in Antiferromagnetic Substances. *Sov. Phys. JETP* **1957**, 5 (6), 1259–1272.

(33) Moriya, T. Anisotropic Superexchange Interaction and Weak Ferromagnetism. *Phys. Rev.* **1960**, 120 (1),

91–98.

(34) Bogdanov A. N.; Yablonskii, D. A. Thermodynamically Stable "Vortices" in Magnetically Ordered Crystals. The Mixed State of Magnets. *Sov. Phys. JETP* **1989**, 68 (1), 101.

(35) Cortés-Ortuño, D.; Beg, M.; Nehruji, V.; Breth, L.; Pepper, R.; Kluyver, T.; Downing, G.; Hesjedal, T.; Hatton, P.; Lancaster, T.; Hertel, R.; Hovorka, O.; Fangohr, H. Proposal for a Micromagnetic Standard Problem for Materials with Dzyaloshinskii-Moriya Interaction. arXiv:1803.11174

(36) The DMI energy $E_{DM}$ for the bulk- and D2d-type DMIs read: $E_{DM} = \int D\mathbf{m} \cdot (\nabla \times \mathbf{m}) \, \mathrm{d}^3 \boldsymbol{r}$ and $E_{DM} = \int D\mathbf{m} \cdot (\partial \mathbf{m}/\partial x \times \hat{\mathbf{e}}_x - \partial \mathbf{m}/\partial y \times \hat{\mathbf{e}}_y) \, \mathrm{d}^3 \boldsymbol{r}$, respectively.

(37) Zhang, S. L.; van der Laan, G.; Wang, W. W.; Haghighirad, A. A.; Hesjedal, T. Direct Observation of Twisted Surface Skyrmions in Bulk Crystals. *Phys. Rev. Lett.* **2018**, 120 (22), 227202.

(38) Donahue M. J.; Porter, D. G. OOMMF User's Guide Version 1.0 Interagency Report NISTIR 6376, National Institute of Standards and Technology: Gaitherburg, MD, (1999). http://math.nist.gov/oommf/

(39) The excitation field with a radial shape [Fig. 1(e)] can be generated by a thin toroidal antenna, as shown in Fig. S1. In this theoretical study, the antenna is factitiously embedded into the prism without breaking the sample for a direct comparison of the oppositely propagating spin waves. In a realistic device, such an insertion is impossible unless the prism is cut into two pieces. Thus, in an experimental study, one can place the antenna on top of a square surface of the prism.

(40) We note that $\alpha = 0.01$ is a very low value for known DMI materials. Choosing such a low value in a computation study (8-10,12) enables strong excitation and long-distance propagation of spin waves, without modifying the key physics, and therefore is favorable for the identification of the mode structures of spin waves. The pioneering theoretical works (8,9,12) of domain wall-based magnonic waveguides also assumed low damping values for DMI materials.

(41) Berkov D. V.; Gorn, N. L. Micromagnetic Simulations of the Magnetization Precession Induced by a Spin-Polarized Current in a Point-Contact Geometry (Invited). *J. Appl. Phys.* **2006**, 99 (8), 08Q701.

(42) Dvornik, M.; Au, Y.; Kruglyak, V. V. Magnonics: Micromagnetic Simulations in Magnonics. *Top. Appl. Phys.* **2013**, 125, 101–115.

(43) García-Cervera, C. J.; Gimbutas, Z.; Weinan, E. Accurate Numerical Methods for Micromagnetics Simulations with General Geometries. *J. Comput. Phys.* **2003**, 184 (1), 37–52.

(44) Garcia-Sanchez, F.; Borys, P.; Vansteenkiste, A.; Kim, J. V.; Stamps, R. L. Nonreciprocal Spin-Wave Channeling along Textures Driven by the Dzyaloshinskii-Moriya Interaction. *Phys. Rev. B* **2014**, 89 (22), 224408.

(45) Hoffmann, M.; Zimmermann, B.; Müller, G. P.; Schürhoff, D.; Kiselev, N. S.; Melcher, C.; Blügel, S. Antiskyrmions Stabilized at Interfaces by Anisotropic Dzyaloshinskii-Moriya Interactions. *Nat. Commun.* **2017**, 8, 308.

(46) Jorzick, J.; Demokritov, S.; Hillebrands, B.; Bailleul, M.; Fermon, C.; Guslienko, K.; Slavin, A.; Berkov, D.; Gorn, N. Spin Wave Wells in Nonellipsoidal Micrometer Size Magnetic Elements. *Phys. Rev. Lett.* **2002**, 88 (4), 047204.

(47) Demidov, V. E.; Demokritov, S. O.; Rott, K.; Krzysteczko, P.; Reiss, G. Mode Interference and Periodic Self-Focusing of Spin Waves in Permalloy Microstripes. *Phys. Rev. B* **2008**, 77 (6), 064406.

(48) Xing, X.; Li, S.; Huang, X.; Wang, Z. Engineering Spin-Wave Channels in Submicrometer Magnonic Waveguides. *AIP Adv.* **2013**, 3, 032144.

(49) Takagi, R.; Morikawa, D.; Karube, K.; Kanazawa, N.; Shibata, K.; Tatara, G.; Tokunaga, Y.; Arima, T.; Taguchi, Y.; Tokura, Y.; Seki, S. Spin-Wave Spectroscopy of the Dzyaloshinskii-Moriya Interaction in Room-Temperature Chiral Magnets Hosting Skyrmions. *Phys. Rev. B* **2017**, 95 (22), 220406.

(50) Manna, K.; Sun, Y.; Muechler, L.; Kübler, J.; Felser, C. Heusler, Weyl and Berry. *Nat. Rev. Mater.* **2018**, 3 (8), 244–256.

(51) Gaier, O.; Hamrle, J.; Trudel, S.; Parra, A. C.; Hillebrands, B.; Arbelo, E.; Herbort, C.; Jourdan, M. Brillouin Light Scattering Study of $Co_2Cr_{0.6}Fe_{0.4}Al$ and $Co_2FeAl$ Heusler Compounds. *J. Phys. D: Appl. Phys.* **2009**, 42 (8), 084004.

(52) Rana B.; Otani, Y. Voltage-Controlled Reconfigurable Spin-Wave Nanochannels and Logic Devices. *Phys. Rev. Appl.* **2018**, 9 (1), 014033.

(53) Zheng, F.; Rybakov, F. N.; Borisov, A. B.; Song, D.; Wang, S.; Li, Z.-A.; Du, H.; Kiselev, N. S.; Caron, J.; Kovács, A.; Tian, M.; Zhang, Y.; Blügel, S.; Dunin-Borkowski, R. E. Experimental Observation of Chiral Magnetic Bobbers in B20-Type FeGe. *Nat. Nanotechnol.* **2018**, 13 (6), 451–455.

(54) Vogt, K.; Fradin, F.; Pearson, J.; Sebastian, T.; Bader, S.; Hillebrands, B.; Hoffmann, A.; Schultheiss, H. Realization of a Spin-Wave Multiplexer. *Nat. Commun.* **2014**, 5, 3727.

(55) Milde, P.; Köhler, D.; Seidel, J.; Eng, L.; Bauer, A.; Chacon, A.; Kindervater, J.; Mühlbauer, S.; Pfleiderer, C.; Buhrandt, S. Unwinding of a Skyrmion Lattice by Magnetic Monopoles. *Science* **2013**, 340 (6136), 1076–1080.

**FIGURE CAPTIONS**

**FIG. 1** Equilibrium configurations of magnetization in a square prism and the excitation field distribution. (a) AntiSkyrmion tube and (b) quasiuniform single domain, formed in the prism, with D2d-type DMI. (c) Skyrmion tube and (d) quasiuniform single domain, in the prism, with bulk-type DMI. (e) Spatial profile of the excitation field on the middle cross-sectional plane of the prism.

**FIG. 2** Magnon dispersion along various lines in the prism. The static spin configurations in (a) and (b) are antiSkyrmion tube and quasiuniform single domain, respectively. All the lines 1–5 are parallel to the long axis ($z$-axis) of the prism and intersect the cross-sectional plane ($xy$-plane) at points 1–5, respectively, as denoted in the inset.

**FIG. 3** Magnon propagation patterns and mode profiles in the prism. Both Fourier amplitude and phase components of $m_z$ are shown in each of (a)–(d). The propagation patterns are along the rectangular cut-planes at $y = x$, $x = w/2$, and $y = w/2$, while the mode profiles are from the cross-sectional planes at $z = l/4$, $z = l/2$, and $z = 3l/4$. [(a) and (c)] AntiSkyrmion-tube and [(b) and (d)] quasiuniform single-domain states are assumed as the magnetic background. The magnon frequencies are 19.76 GHz [(a) and (b)] and 34.58 GHz [(c) and (d)], respectively.

**FIG. 4** Magnon dispersion along various lines in the prism. The static spin configurations in (a) and (b) are Skyrmion tube and quasiuniform single domain, respectively. All the lines 1–5 are parallel to the long axis ($z$-axis) of the prism and intersect the cross-sectional plane ($xy$-plane) at points 1–5, respectively, as denoted in the inset.

**FIG. 5** Magnon propagation patterns and mode profiles in the prism. Both Fourier amplitude and phase components of $m_z$ are shown in each of (a)–(d). The propagation patterns are along the rectangular cut-planes at $y = x$, $x = w/2$, and $y = w/2$, while the mode profiles are from the cross-sectional planes at $z = l/4$, $z = l/2$, and $z = 3l/4$. [(a) and (c)] Skyrmion-tube and [(b) and (d)] quasiuniform single-domain states are assumed as the magnetic background. The magnon frequencies are 10.87 GHz [(a) and (b)] and 50.39 GHz [(c) and (d)], respectively.

**FIG. 6** Distribution of internal magnetic field over the cross-sectional plane of the prism. AntiSkyrmion-tube and quasiuniform single-domain states, with D2d-type DMI, are assumed as the magnetic background in (a) and (b), respectively. Skyrmion-tube and quasiuniform single-domain states, with bulk-type DMI, are assumed as the magnetic background in (c) and (d), respectively.

**FIG. 7** Magnetic field control of the magnon dispersion in the square prism. [(a) and (b)] AntiSkyrmion tube and [(c) and (d)] Skyrmion tube are assumed as the magnetic background, respectively. From column #1 to #3, the magnetic field $H_{bias}$ is increased, as denoted in each panel. Magnon dispersion along the propagation routes line 1 [(a) and (c)] and line 4 [(b) and (d)] are considered.

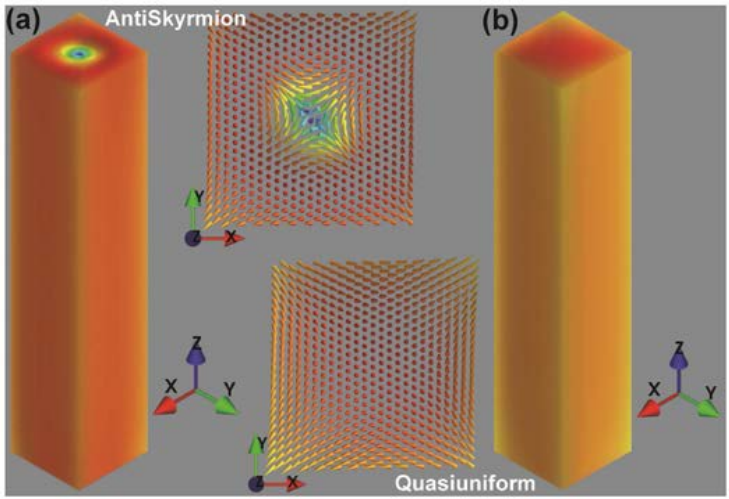

**(a)** AntiSkyrmion

**(b)**

Quasiuniform

**(c)** Skyrmion

**(d)**

Quasiuniform

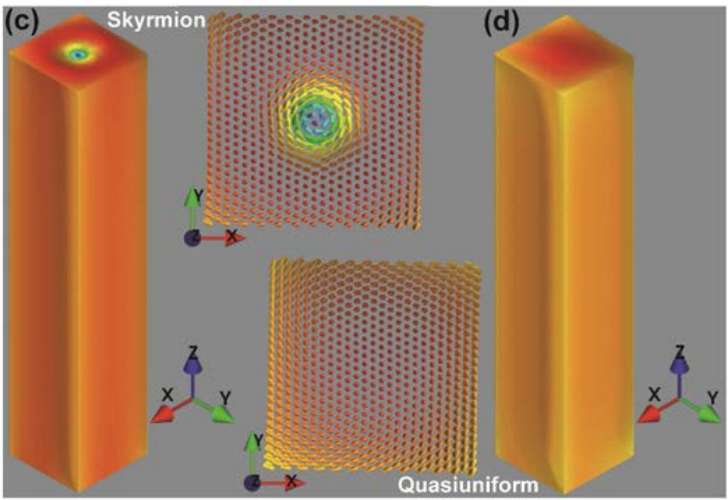

**(e)** Excitation

$h(t)$

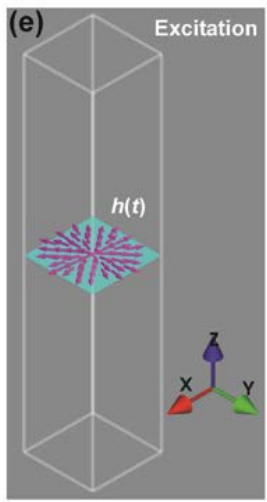

**(a)**

**(b)**

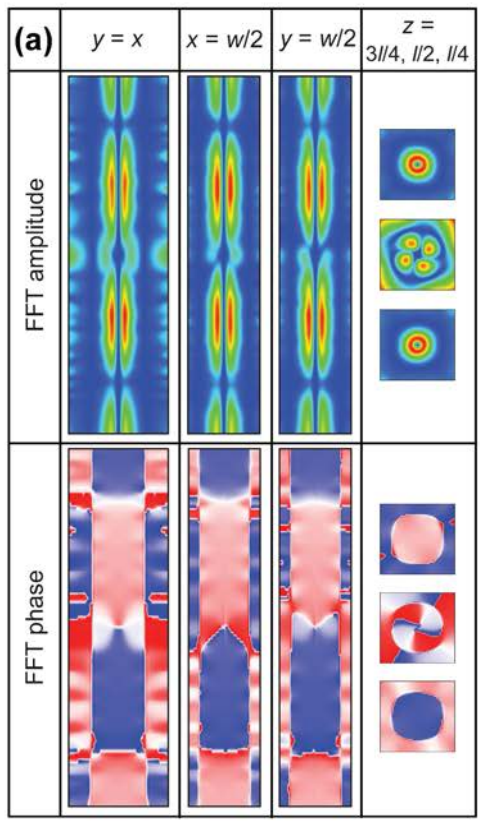

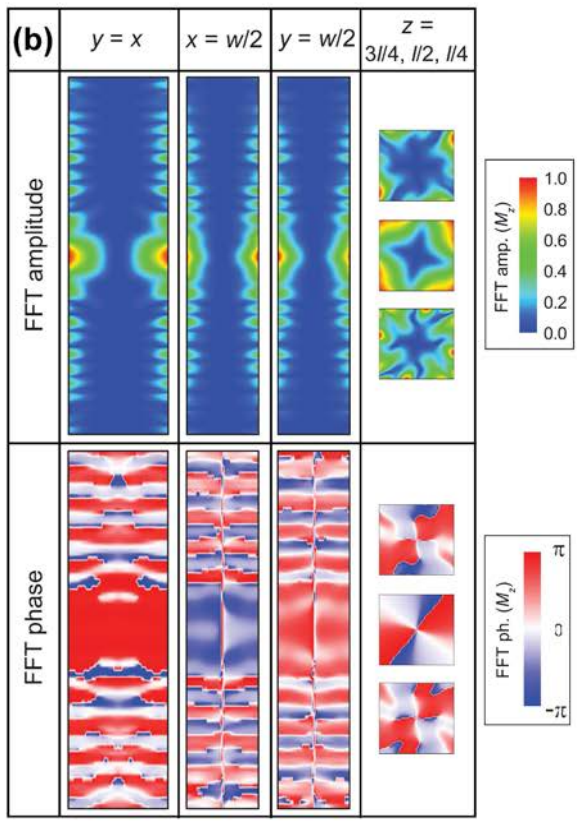

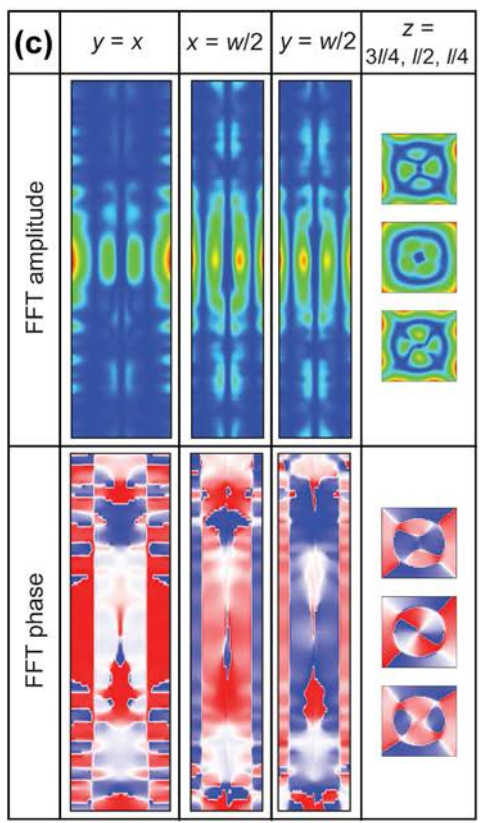

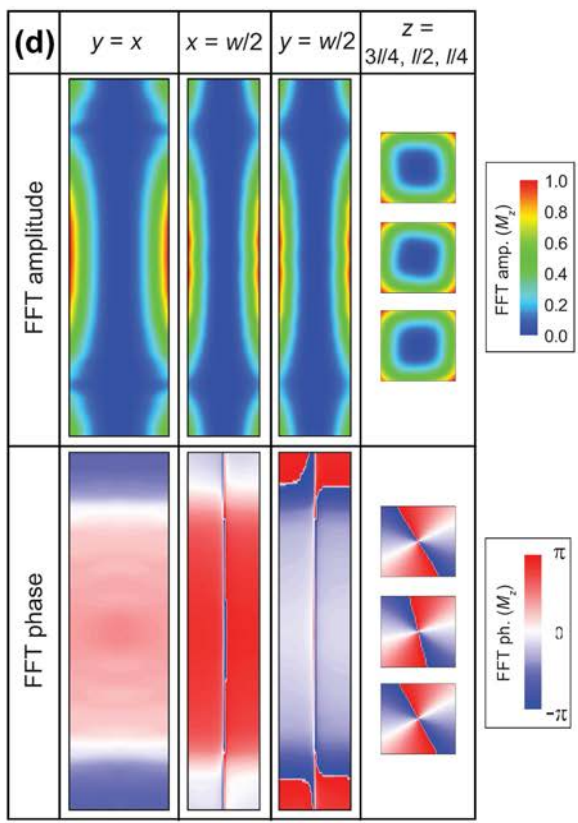

(a)

#1 | #2 | #3 | #4 | #5

Skyrmion Line 1, Line 2, Line 3, Line 4, Line 5

(b)

Quasiuniform Line 1, Line 2, Line 3, Line 4, Line 5

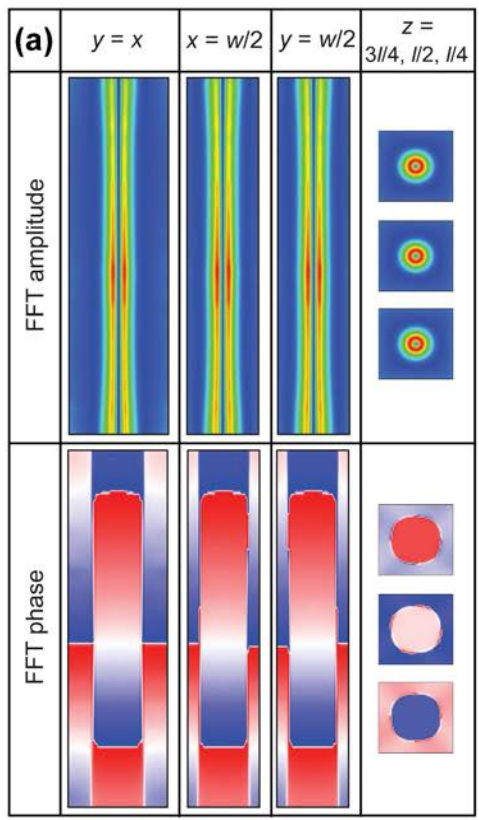

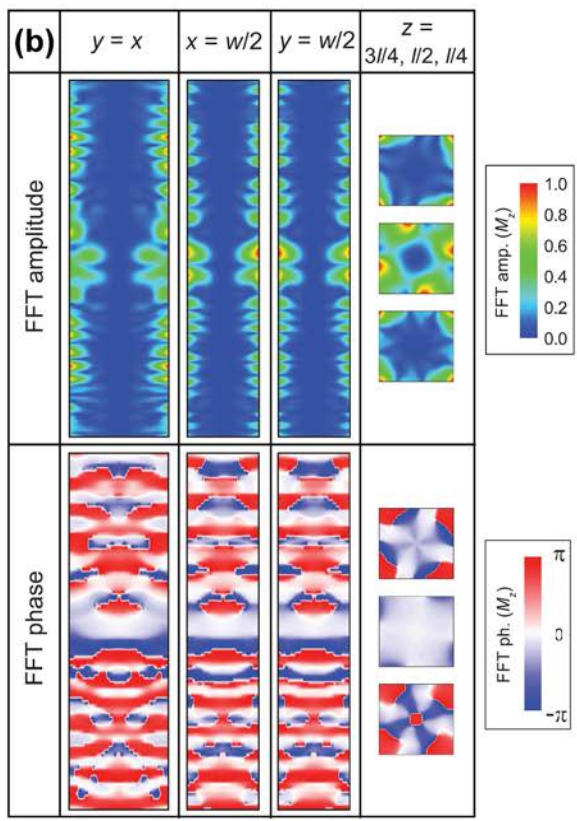

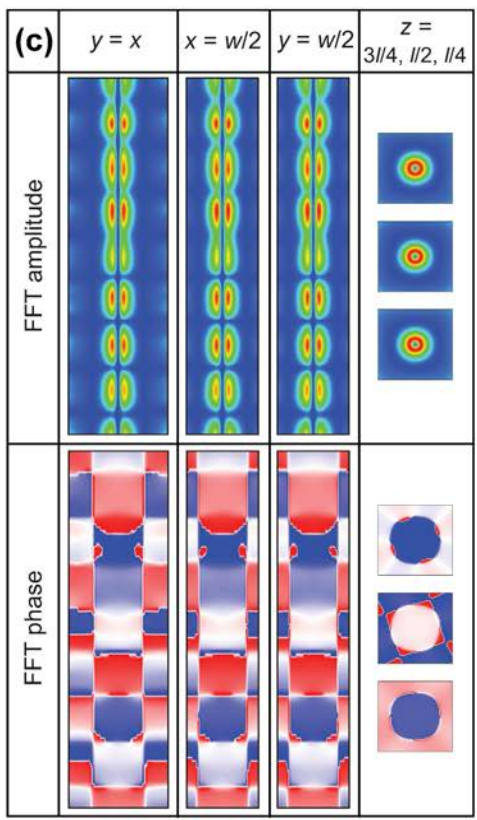

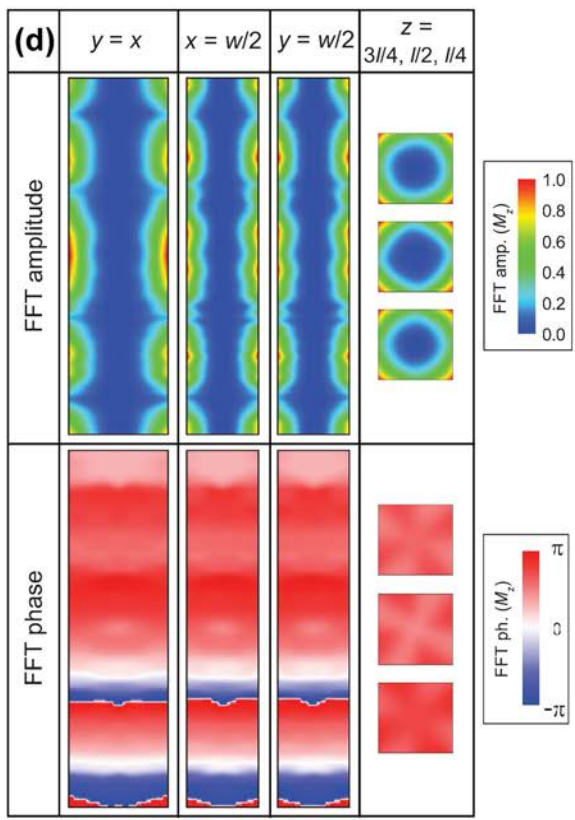

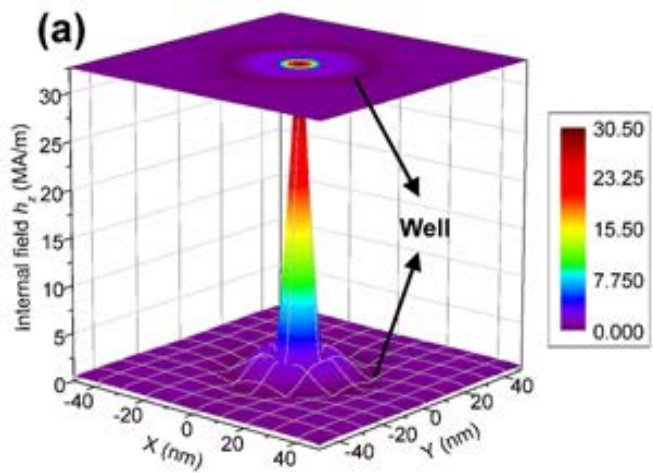

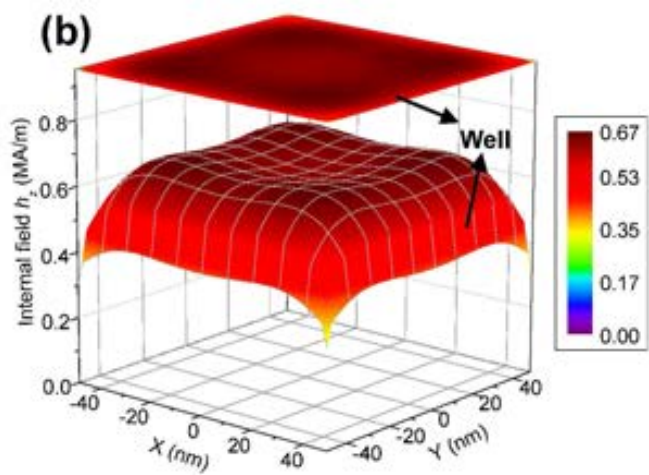

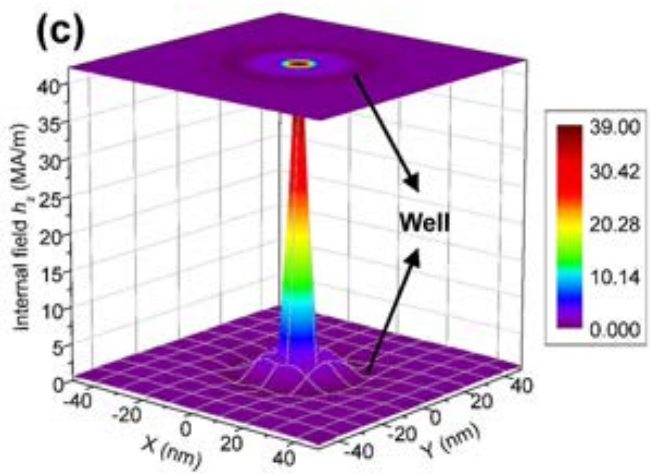

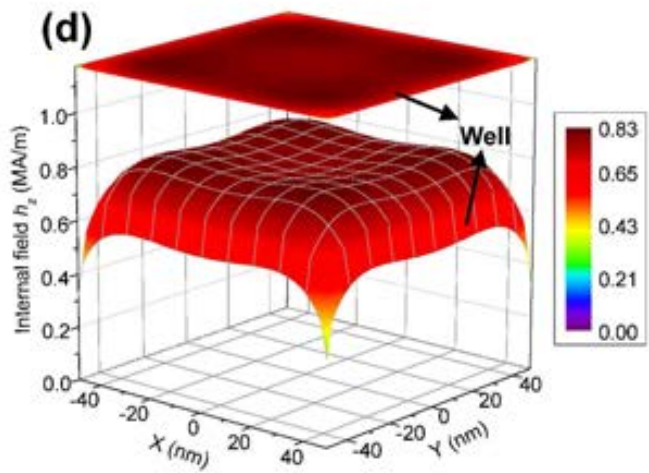

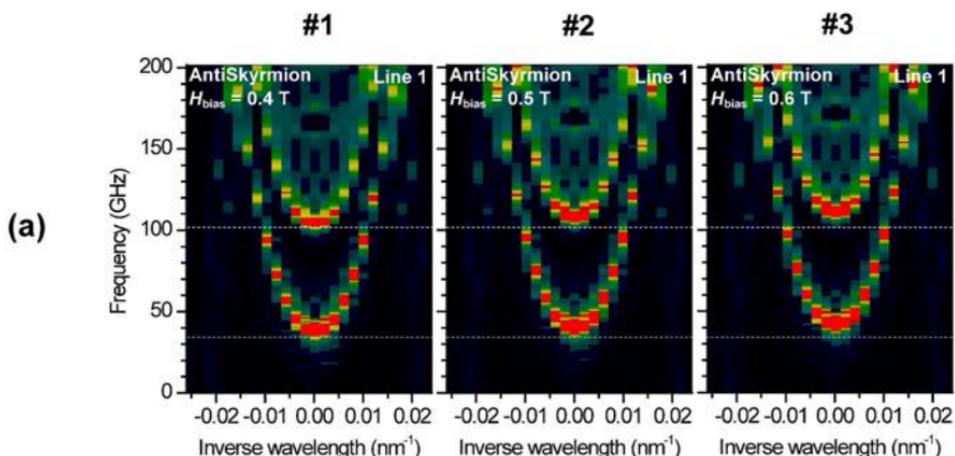

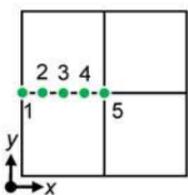

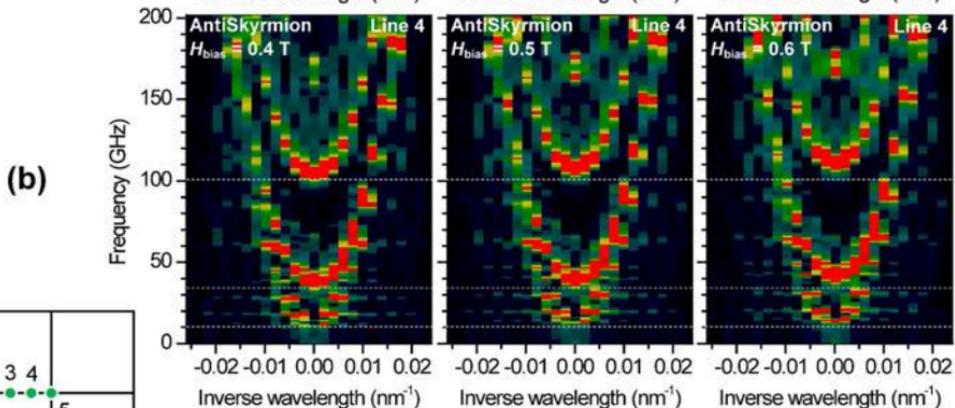

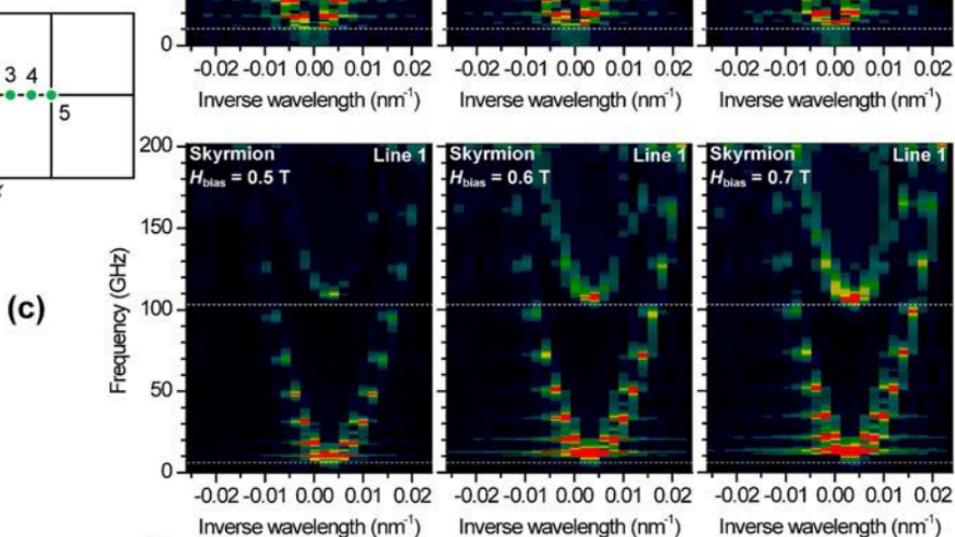

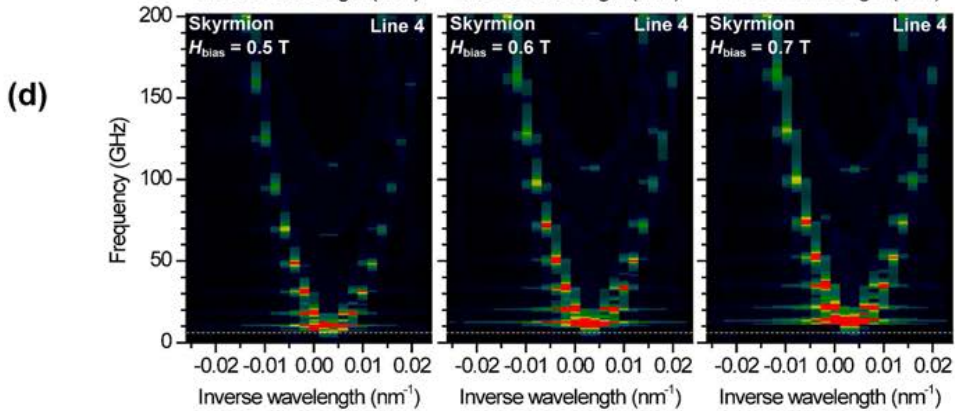

SUPPORTING INFORMATION FOR

# Skyrmion Tubes as Magnonic Waveguides

Xiangjun Xing,*[,†] Yan Zhou,*[,‡] and H. B. Braun*[,§]

[†]*School of Physics and Optoelectronic Engineering, Guangdong University of Technology, Guangzhou 510006, China*

[‡]*School of Science and Engineering, The Chinese University of Hong Kong, Shenzhen, 518172, China*

[§]*School of Physics, University College Dublin, Dublin 4, Ireland*

*Email: xjxing@gdut.edu.cn; zhouyan@cuhk.edu.cn; beni.braun@ucd.ie

FIG. S1

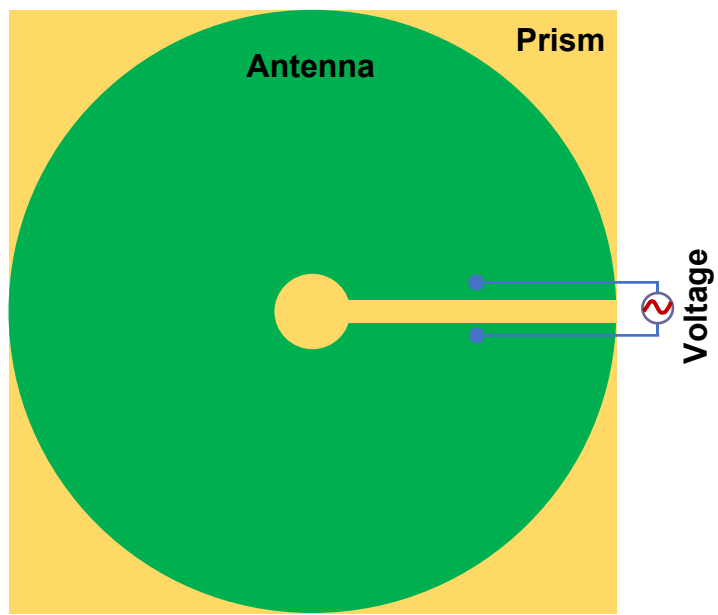

FIG. S1 A thin toroidal antenna that can creates the excitation field with a radial shape as shown in Fig. 1(e).

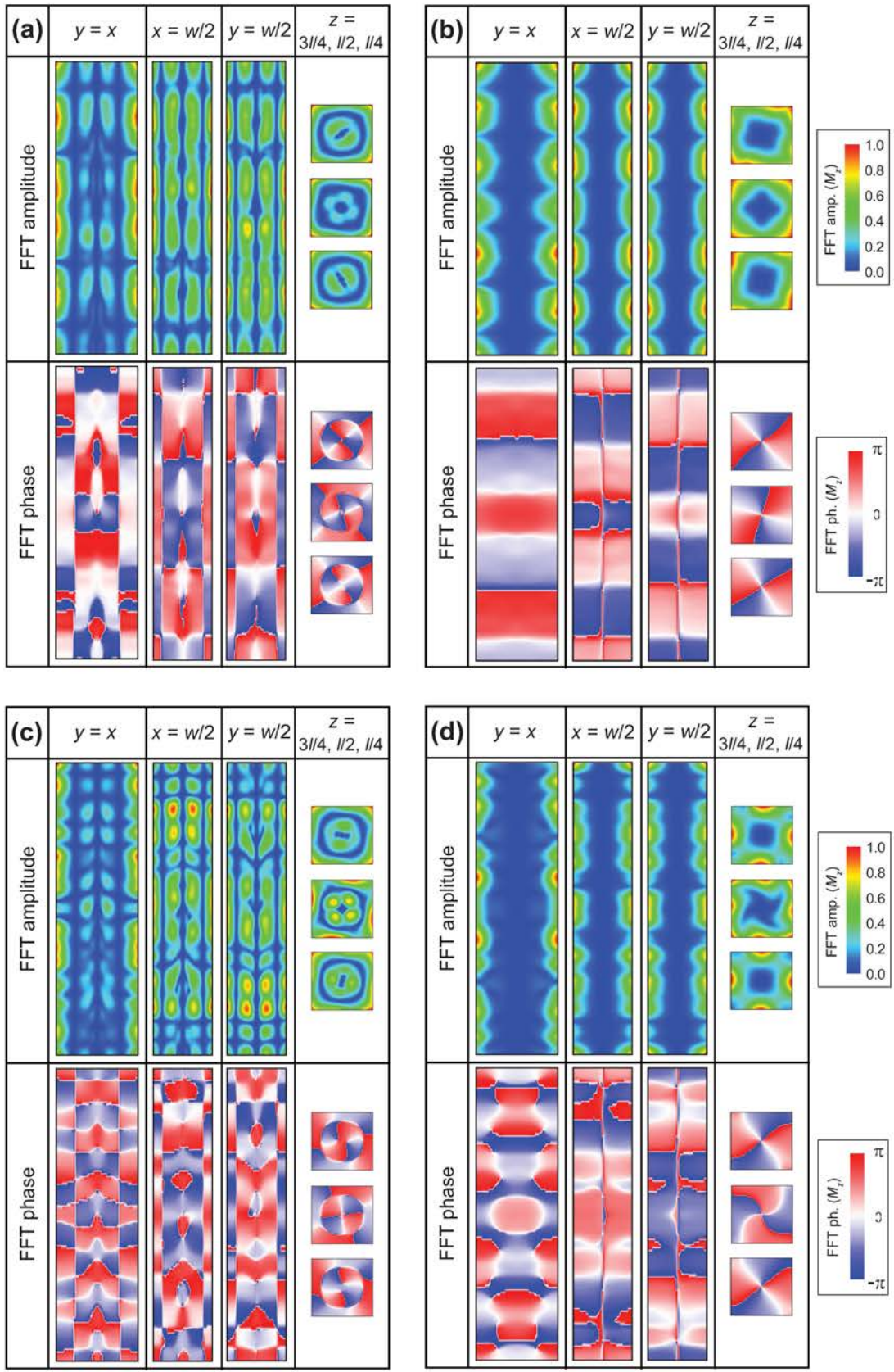

FIG. S2 Magnon propagation patterns and mode profiles in the prism. Both Fourier amplitude and phase components of $m_z$ are shown in each of (a)–(d). The propagation patterns are along the rectangular cut-planes at $y = x$, $x = w/2$, and $y = w/2$, while the mode profiles are from the cross-sectional planes at $z = l/4$, $z = l/2$, and $z = 3l/4$. [(a) and (c)] AntiSkyrmion-tube and [(b) and (d)] quasiuniform single-domain states are assumed as the magnetic background. The magnon frequencies are 50.39 GHz [(a) and (b)] and 119.54 GHz [(c) and (d)], respectively.

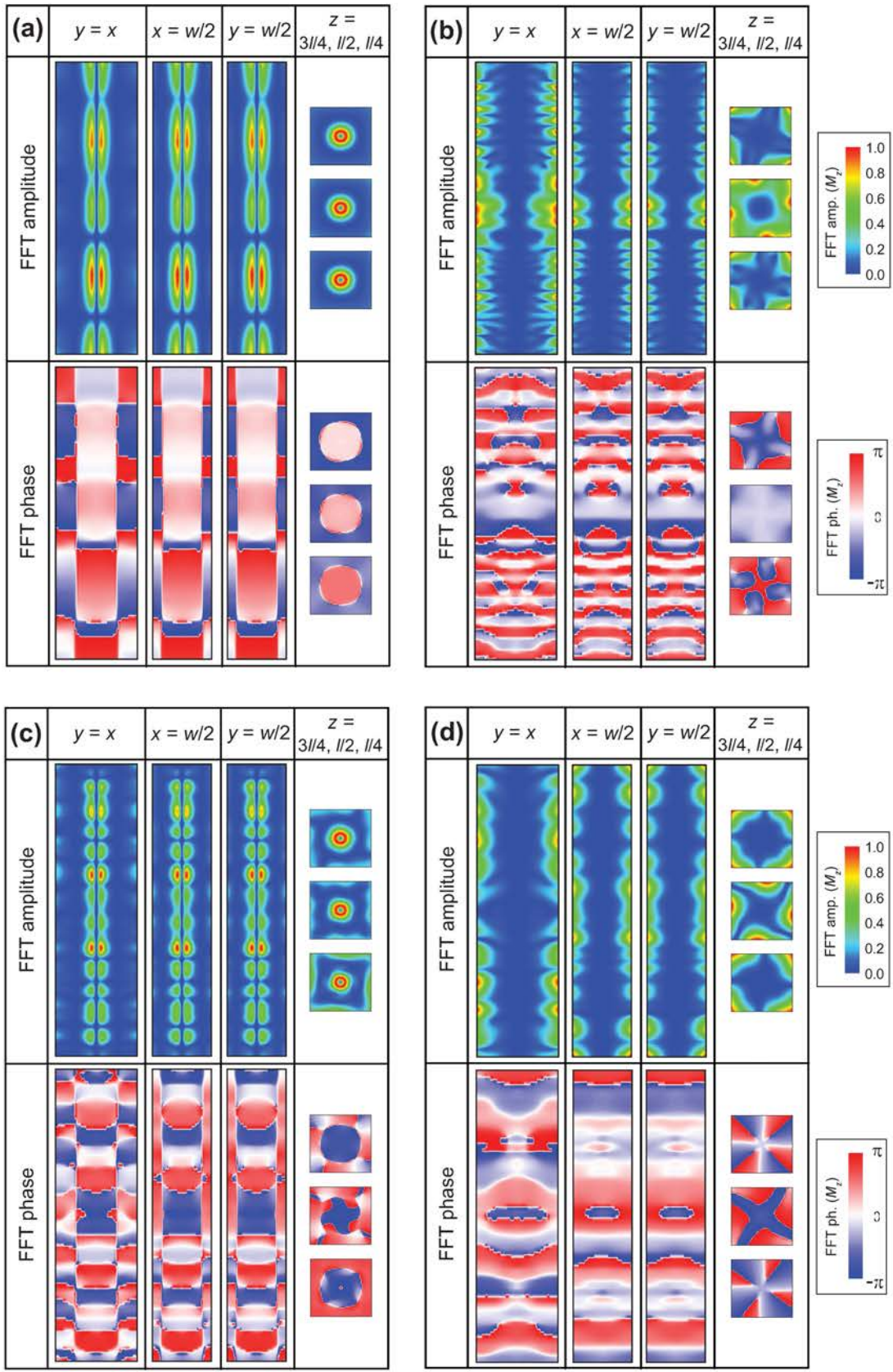

FIG. S3 Magnon propagation patterns and mode profiles in the prism. Both Fourier amplitude and phase components of $m_z$ are shown in each of (a)–(d). The propagation patterns are along the rectangular cut-planes at $y = x$, $x = w/2$, and $y = w/2$, while the mode profiles are from the cross-sectional planes at $z = l/4$, $z = l/2$, and $z = 3l/4$. [(a) and (c)] Skyrmion-tube and [(b) and (d)] quasiuniform single-domain states are assumed as the magnetic background. The magnon frequencies are 19.76 GHz [(a) and (b)] and 119.54 GHz [(c) and (d)], respectively.